\documentstyle[12pt]{article}
\def\bb{\begin{eqnarray}}
\def\ee{\end{eqnarray}}
\def\eee{\nonumber\end{eqnarray}}

\newcommand {\spartial}{\partial \hspace{-0.21cm}\slash}
\newcommand{\B}{\hspace{-0.1cm}}

\newcommand{\nc}{non-commutative }
\newcommand{\Nc}{Non-commutative }
\newcommand{\ac}{\mbox{$\cal A$ }}

\newcommand{\PD}{Poincar\'{e} Duality }

\newcommand{\hil}{\mbox{$\cal H$ }}
\newcommand{\mata}{\rho_w(a,b,c)=
\bordermatrix{&  {L} &  {R}
\cr
       & \rho_1(a)&  \cr
     & & \rho_2(b)  \cr}}
\newcommand{\mataa}{\mbox{with}\;\; \rho_1(a)=
\bordermatrix{&  {(u,d)_L} &  {(\nu,e)_L}
\cr
       & a\otimes 1_3&  \cr
     & & a  \cr}
\;\;\mbox{and}\;\; \rho_2(b)=
\bordermatrix{&  {(u,d)_R} &  {(\nu,e)_R}
\cr
       & B\otimes 1_3&  \cr
     & & B  \cr}}

\begin{document}
\hsize 17truecm
\vsize 24truecm
\font\twelve=cmbx10 at 13pt
\font\eightrm=cmr8
\baselineskip 18pt

\begin{titlepage}
\centerline{\twelve DEPARTMENT OF MATHEMATICAL SCIENCES}
\centerline{\twelve Durham University}
\centerline{\twelve Durham DH1 3LE}
\centerline{\twelve England}
\vspace{0.5cm}
\centerline{\twelve CENTRE DE PHYSIQUE THEORIQUE}
\centerline{\twelve CNRS - Luminy, Case 907}
\centerline{\twelve 13288 Marseille Cedex}
\centerline{\twelve France}
\vskip 4truecm

\centerline{\twelve Non-Commutative Geometry and the Strong Force}

\bigskip

\begin{center}
{\bf Becca Asquith}
\footnote{funded by PPARC \newline
r.e.asquith@dur.ac.uk}
\end{center}

\vskip 1truecm
\leftskip=1cm
\rightskip=1cm
\centerline{\bf Abstract}

\medskip

\noindent The restrictions imposed on the strong force in the `\nc
standard  model' are examined. It is concluded that given the
framework of non-commutative geometry and assuming the electroweak sector
of the standard model many details of the strong force can be explained
including its vectorial nature.

\vskip 0.5truecm
PACS-92: 11.15 Gauge field theories\\
\indent
MSC-91: 81E13 Yang-Mills and other gauge theories

\vskip 2truecm

\noindent sept 1995
\vskip 1truecm
\noindent DTP/95/49\\
\noindent CPT-95/P.3239\\

\vskip1truecm

 \end{titlepage}

\newpage

\section{Introduction}

The standard model (SM) is extremely successful at predicting
experimental  results but from a theoretic/aesthetic point of view it is
much  less satisfactory. It contains many arbitary inputs, in particular:
\begin{enumerate}
\item \underline{The Higgs Sector.} Whilst the Yang-Mills sector is well
motivated geometrically this results in massless particles so the Higgs
sector  has to be tagged on at the end by hand.
\item \underline{Gauge Group and Group Representation.} The choice of
gauge  group is arbitary and given a gauge group the choice of
irreducible  representation from the infinite number available is also
arbitary.
\item \underline{Parity structure and Spontaneous Symmetry Breaking.} In the
SM the SU(2) weak force is assumed to be maximally parity violating and to be
 spontaneously broken whilst the SU(3) force is assumed to be vectorial and to
 remain unbroken. In the
SM there is no reason for this apparent link between parity structure
and  spontaneous symmetry breaking. \end{enumerate}

\noindent Non-commutative geometry goes a long way towards answering some
of  these problems. This paper does not contain the (by now well documented)
details of  \nc geometry. For an extremely thorough mathematical explanation
of \nc geometry please see
\cite{cb} or \cite{th}, for a clear explanation more suited to the physicist
please see \cite{sz}. \Nc geometry is a generalization of the tools of
classical geometry (eg the differential calculus, the notion of a metric
space) to a much wider class of manifolds -the collection of `\nc manifolds'.
Since there is a differential calculus on these manifolds it is possible
to  define a Yang-Mills action over them \cite{cl}. This is precisely the area
 that is of interest to particle physicists. If a manifold
$X=M_4 \times \{0,1\}$ ie two copies of a Riemannian
manifold separated by a finite distance, is considered (a manifold that cannot
 be treated with classical geometry due to the discreteness of the space) and
a pure Yang-Mills theory is constructed over this manifold then something very
 interesting happens.
 A `gauge boson' associated with the discreteness of the space occurs. This
gauge boson has spin zero and a quartic potential of the form required for
spontaneous symmetry breaking. So a pure Yang-Mills theory over such a space
 automatically has a natural
 Higgs sector. Furthermore this Higgs sector only arises if the representation
 of the algebra (associated with the forces -please see below for details) on
the Hilbert space of left handed fermions is different to that of the
 representation on the space of right handed fermion. That is there is a clear
 explanation for the correlation between spontaneous symmetry breaking and
parity structure that is observed in the standard model (as noted in point 3
 above). `\Nc standard models' also help to solve problem
 2. The choice of gauge group is still more or less arbitary (though the
 exceptional groups are ruled out -something that is not the case in the usual
 formulation of the standard model).
 However the choice of representations of the gauge group is greatly
restricted. The representation of the group comes from the representation of
 the algebra -this is a very restrictive condition. In general a group has an
 infinite number of unitary
irreducible representations whereas an algebra has typically one or two.
 See \cite{isym} for a detailed analysis of this subject. \Nc geometry
 can therefore be seen to motivate many of the previously arbitary features of
 the SM.

The topic of this paper is another constraint imposed on the SM by \nc
geometry, namely that given the structure of the electroweak sector it follows
 automatically that the strong force is vectorial (ie non-chiral) and
therefore that SU(3) is not broken.
This is an explicit proof of an idea of Alain Connes.
\section{Basic Framework of Non-Commutative Geometry}
Non-commutative Yang-Mills models are constructed via a K-cycle
$(\ac\B,\hil\B,D)$ with a real structure J.
\ac is an involutive algebra whose group of unitaries is the gauge group of
the model.
\hil is the Hilbert space of the fermions on which \ac is represented by
 $\rho$ and D is the generalized Dirac operator.
D enables us to define a metric in the non-commutative setting.
The real structure J is the non-commutative generalization of the charge
conjugation operator \cite{c6}.
It satisfies the following conditions:
\begin{enumerate}
\item $JD=DJ$
\item $J^2=\pm 1$
\item $[\rho(y),J\rho(y')J^{-1}]=0\;\;\;\;  y,y'\in \ac$
\item $[[D,\rho(y)],J\rho(y')J^{-1}] =0 \;\;\;\; y,y' \in \ac$
\end{enumerate}

\noindent Condition 4, an important condition in the calculations in this
 paper can also be arrived at by considering \PD\B.
Classically all manifolds have an isomorphism known as \PD\B\B, this is not
 the case for \nc manifolds where \PD has to be imposed.
The conditions for the existence of such an isomorphism are:
\begin{enumerate}
\item $ [[D,y],Jy'J^{-1}]=0\;\;\;  y,y'\in \ac $
\item $Tr_{\omega} (\gamma [D,y^1]...[D,y^n] \!\mid\!\! D\! \mid ^{-n})
=0 \;\;\; y^j \in \cal A $
\end{enumerate}
Note that condition 1 immediately above is the same as condition 4 on J.
 For an explanation of condition 2 (not needed for this paper) see
 \cite{cbp}.
For the \nc standard model the K-cycle $(\ac\B,\hil\B,D)$ is taken:
\[\ac =C^{\infty }(M)\otimes [H \oplus C \oplus M_3(C) ]\]
\[D=\spartial \otimes 1 + \gamma_5 \otimes D_f\]
where H are the quaternions, $D_f$ is the fermionic mass matrix and \hil the
Hilbert space of left and right fermions and anti-fermions.
Note that in this case it is possible to split the space into a finite part
 and an infinite part. The finite part of the algebra and the generalized
Dirac operator ($\ac_f=H \oplus C \oplus M_3(C)$ and $D_f$ respectively) are
 represented on the finite space. In
 the calculations that follow just the finite sector is worked with, the full
 model is then obtained by tensoring with the infinite sector.
For details and predictions of the \nc standard model see \cite{isfm}.

\section{Calculations}
Aim: the aim of these calculations is to show that within the framework of \nc
 geometry, if the electroweak sector is assumed the strong force is
 constrained to be vectorial.

\noindent The assumptions of this calculation are then

\[[\rho(y),J\rho(y')J^{-1}]=0\;\;\;\;  y,y'\in \ac\]
\[[[D,\rho(y)],J\rho(y')J^{-1}] =0 \;\;\;\; y,y' \in \ac\]
these are as discussed a direct consequence of the framework of
 non-commutative geometry (conditions 3 and 4 in the previous section).
 Further the form of the electroweak sector is assumed, ie the algebra \ac is
 taken to be
\[\ac=H \oplus C \oplus X\]
where X (the algebra associated with the strong force) is assumed to be a
simple algebra. The representation of \ac on \hil is given by
\[\rho(y) =\left[\begin{array}{cc}
                \rho_w(y) &  \\
                       & \rho_s(y)
        \end{array}\right]\;\;\;y\in\ac \]
where $y=(a,b,c) \in H \oplus C \oplus X$
\[\mata \]
\vspace{0.3cm}
  \[\mataa ,\;\; B=\left[\begin{array}{cc}
                                                    b &  \\
                                                   &  \bar b
                                            \end{array} \right].\]

\vspace{0.2cm}
\noindent From experimental evidence \cite{hm} it is known that quarks exist
 in `threes' (ie what we call colour triplets) of identical mass so the form
of the fermionic mass matrix is known. The following notation is used
\[D_f=\left[\begin{array}{cccc}
                0 & M & 0 & 0  \\
               M^* &0 & 0 & 0  \\
                0 & 0 & 0 & M \\
                0 & 0 & M^* & 0
    \end{array}\right]
\mbox{where}\;\;\;
M=\left[\begin{array}{cc}
              M_q\otimes1_3 &  \\
               & M_l
     \end{array}\right],\]
\[M_q=\left[\begin{array}{cc}
              m_u &  \\
               & m_d
     \end{array}\right],
\;\;\;M_l=\left[\begin{array}{cc}
              m_\nu &  \\
               & m_e
     \end{array}\right].\]

 \noindent In summary then the only fact that is assumed about the strong
force is that its
associated algebra is simple.
\vspace{0.3cm}
\newline \noindent Summary of calculations:

\noindent From $[y,Jy'J^{-1}]=0$ it follows that $[\rho_w,\rho_s]=0$ so
 $\rho_s$ is block diagonal. Also since $\rho_s$ is an algebra representation
 it cannot depend on $a$ if it is to commute with $\rho_w$.
So
\[\rho_s(b,c)=\left[\begin{array}{cc}
                      \rho_3(b,c) & \\
                            &  \rho_4(b,c)
              \end{array}\right]\]
the condition $[\rho_w,\rho_s]=0$ is then equivalent to
\[ [\rho_1,\rho_3]=[\rho_2,\rho_4]=0\;\;\;\;(i).\]
Now consider the second condition
$ [[D,y],Jy'J^{-1}]=0$.
Given (i) it is trivial matrix multiplication to show that this second
condition is equivalent to
\[M\rho_2\rho_4 -\rho_1 M \rho_4 -\rho_3 M \rho_2 + \rho_3\rho_1 M =0.\]

\noindent Now only $\rho_1$ depends on $a$ so this equation is actually two
equations which must be satisfied separately
\[M\rho_2\rho_4 -\rho_3 M \rho_2 =0\]
and
\[-\rho_1 M \rho_4+ \rho_3\rho_1 M =0.\]
These equations contain essentially the same information so consider only the
latter. If $a$ is taken to be 1 then since $\rho_1$ is a representation we
have $\rho_1(1)=1$
\[
\rho_3M=M\rho_4\;\;\;\;(ii).\]
Given (i) and Schur's lemma it follows that as a matrix
\[\rho_3=1_2\otimes R \oplus  1_2K\;\;\;\;R\in M_3(C),K \in C\]
furthermore it is known that $\rho_3$ is a representation of C and X
 therefore $R=c\;\; \mbox{or}\;\; \bar{c}$ and
 $K=b\;\; \mbox{or}\;\; \bar{b}$. Similarly it can be seen that
\[\rho_4(b,c)=1_2\otimes T \oplus \Delta \]
where $T=c \;\;\mbox{or}\;\;\bar c$ and $\Delta =
\left[\begin{array}{cc}
      b   &   \\
          & b
 \end{array}\right],
\left[\begin{array}{cc}
      \bar b   &   \\
          & b
 \end{array}\right],
\left[\begin{array}{cc}
      b   &   \\
          & \bar b
 \end{array}\right]\;\;\mbox{or}\;\;
\left[\begin{array}{cc}
      \bar b   &   \\
          & \bar b
 \end{array}\right].$

\vspace{0.5cm}
\noindent Now imposing condition (ii) forces $R=T$ and $K1_2=\Delta$
\footnote{this latter equality only holds for $m_{\nu}\neq0$ please see below
for details on massless neutrinos}
ie \[\rho_s(b,c)=\left[\begin{array}{cccc}
     1_2\otimes c &     &     &    \\
                  &   b1_2 & &  \\
                  &            & 1_2 \otimes c & \\
                  &            &   &  b1_2
              \end{array}\right]
\mbox{or}\;\;
 \rho_s(b,c)=\left[\begin{array}{cccc}
     1_2\otimes c &     &     &    \\
                  &  \bar b1_2 & &  \\
                  &            & 1_2 \otimes c & \\
                  &            &   & \bar b1_2
              \end{array}\right].\]
Note then that $\rho_3=\rho_4$ and that $\rho_s$ commutes with the fermionic
mass matrix so it has been shown that the strong force is constrained to be
vectorial.
Also the quark doublet $\left[\begin{array}{c} u\\d\end{array}\right]$ is
acted upon by $1_2 \otimes c$ where c must be a $3\times 3$ matrix. So it
follows that
the strong force doesn't see flavour and that
 $X=M_3(C) \,\;\mbox{or}\;\, M_3(R)$ ie that it's gauge group is either U(3)
or a subgroup of U(3).

It can be seen in the calculation above that I have assumed the presence of a
massive right handed neutrino. This is not a necessary assumption,
the right handed neutrino can be projected out at any stage without affecting
the calculations.
However if a massless right handed neutrino is assumed then it cannot be shown
 that the force on the right handed neutrino is the same as that on the left
handed neutrino.
See \cite{gbn} for work on right handed neutrinos in the framework of
 \nc geometry.

Extra generations: The calculations above have been performed for $N_G=1$, the
 generalization to $N_G=3$ follows the same pattern. There is a slight
 complication that, due to the increased size of the Hilbert space a larger
 number of representations $\rho
_3$ and $\rho_4$ satisfy (i), for instance the undesirable
\[\rho_3=(1_2\otimes c \otimes 1_3) \oplus (1_2 \otimes b1_3) \]
in which the strong force mixes the three generations of quarks rather then
the three colours but these are ruled out by (ii).
 So it is found that on the quarks the strong force is again vectorial and
that it is the same on all three generations. The force (associated to C)
on the leptons is again vectorial but is not necessarily the same on every
generation. As for $N_G=1$ if an f-neutrino $(\nu_f, f=e,\mu,\tau)$ is assumed
 to be massless then it cannot be shown that the force on the left f-neutrino
is the same as that on
the right f-neutrino.
\newpage
\section{Conclusions}
So, given the framework of non-commutative geometry and details of the weak
interaction it is possible to predict almost the exact form of the strong
force. In particular that:
\begin{enumerate}
\item the strong force is vectorial
\item the strong force is the same for up as for down quarks
\end{enumerate}
Number (1) is in my opinion the strongest and most suprising prediction and
represents yet another instance in which \nc geometry helps to reduce the
number of arbitary inputs into the standard model. Point (2) follows more
 trivially from the
 fact that $\rho_s$ must be independent of $a \in H$.
\vspace{0.3cm}
\flushleft{\large{Acknowledgement}}

\noindent
I am extremely grateful to David Fairlie, Thomas Sch\"{u}cker and Bruno Iochum
 for their very helpful and constructive comments concerning this work and \nc
 geometry in general.\\
I am also grateful to PPARC for funding this work.
\newpage

\end{document}